# Growth Kinetic studies of ion beam sputtered AlN-thin films: Effect of reactive assistance of nitrogen plasma

Neha Sharma*, K. Prabhakar, S. Ilango, S. Dash and A. K. Tyagi

Material Science Group,

Indira Gandhi Center for Atomic Research,

Kalpakkam, TN, 603102

## 1. Introduction

One of the major needs of thin film device industry has been the deposition of films with smooth and dense surfaces. Ion Beam Sputter Deposition (IBSD) technique facilitates deposition of ultra-smooth thin films with high packing density and excellent film-substrate adhesion. It also has an additional advantage as independent control over fundamental deposition parameters such as ion energy and ion flux can be easily excercised [1-4]. However, it has a major point to disadvantage of slow growth rates which needs to be overcome by assisting the deposition with low energy ions. Low energy ion assistance during IBSD process provides additional activation energy to the growing film and accelerates the nucleation by creating more sites which in turn hastens the growth and subsequent coalescence [5-7]. Hence ion assisted deposition (IAD) alters the kinetics of thin film growth in such a way that the film gets densified with substantially improved structural stability. Considering these facts, it becomes essential to reconceptualize the different growth mechanisms encountered at different stages when ion beam induced sputtering is simultaneously assisted by low energy ions.

Dynamic Scaling Theory (DST) reliably assesses growth mechanism of thin films [8, 9]. This theory propounds that how and up to what extent, the effects of certain surface phenomena like plastic flow, condensation, evaporation and diffusion are altered during the course of low energy ion assisted film growth. These phenomenological informations derived from DST application can quantatively account for enhanced film density and stability. In conventional dynamic scaling theory, scaling behavior is propounded in terms of Family-Vicsek relation [10] described below. We have already applied DST formalism to explain growth of Al films. However, for the sake of completeness, the governing equations are being out lined [11].

$$\xi(L,t) = L^{\alpha} f(t/L^{z}) \qquad (1)$$

Where

$$\xi(L,t) \sim t^{\beta} \quad for \quad t/L^{z} << 1$$

and

$$\xi(L,t) \sim L^{\alpha} \quad for \quad t/L^{z} >> 1$$

Here, ξ is the interface width, L is the length scale over which the roughness is measured, t is the time of growth. α and β are static and dynamic scaling exponents, respectively. z equals α/β.

The autocorrelation function G (|r|) with static scaling exponent 'α' is approximated as:

$$G(|r|) \approx \begin{cases} \delta^2\left[1 - \frac{\alpha+1}{2}\left(\frac{r}{L_c}\right)^{2\alpha}\right], & \text{for } r \le L_c \\ 0, & \text{for } r > L_c \end{cases} \qquad (2)$$

The Fourier transform of equation (2) yields power spectral density function (PSD) g(|q|):

$$g(|q|) \approx \begin{cases} \frac{\alpha}{\pi}\delta^2 L_c^2 & \text{for} \quad |q| < {}^{1}/_{L_c} \\ \frac{\alpha}{\pi}\frac{\delta^2}{L_c^{2\alpha}} q^{-2(\alpha+d)} & \text{for} \quad |q| \geq {}^{1}/_{L_c} \end{cases} \tag{3}$$

where d, in our case, represents line scan direction and equals unity.

To model our data we have used a fitting function similar to one used by William M. Tong et.al. which is expressed as [12]:

$$g(|q|, t) \propto \Omega \frac{\exp(2\sum \chi_n |q|^n t) - 1}{\sum \chi_n |q|^n} \tag{4}$$

Where the q-coefficients $\chi_n$ are simply fitting parameters and n takes the values 1,2,3,4.

The q-coefficients can acquire either positive or negative values which will determine whether the process will be a roughening and smoothening one respectively. According to Herring [13], four normally encountered smoothening mechanisms are plastic flow (n = 1), evaporation-recondensation (n = 2), bulk diffusion (n = 3) and surface diffusion (n = 4).

In this study we have explored the change in appearance and prevalence of certain surface phenomena at each stage of growth of AlN thin films deposited through reactive assistance of $N_2$ plasma. The method of calculation used here is similar to one used in our previous study related to growth kinetics of Al thin films, although no energy or reactive assistance was provided [11]. AlN which is a III-V family compound exhibits synergistic combination of physical, chemical and mechanical properties. These property attributes have made this material indispensable in several technical applications pertaining to thin films, devices and sensors.

In the light of the above implications, we have been motivated to unravel kinetics of reactive ion assisted ion beam sputtered growth of AlN thin films. This constitutes subject matter of present work where in DST formalism has been resorted for dealing with the governing surface related phenomena.

## 2. Experiment

AlN-thin films investigated in this study were grown by IBSD on Si (100) substrate. IBSD system essentially consists of a main deposition chamber and a load lock chamber equipped with a substrate transfer rod. The main deposition chamber is equipped with an RF ion beam source with capability to deliver ions in the energy range 100eV – 2KeV. Then extracted ions are used to sputter the target. An assisted ion source is used concomitantly to assist the sputter deposition process in both reactive and/or non-reactive ways with ions delivered in the range 80 eV to 300 eV. Quartz crystal monitor, residual gas analyzer, substrate heater, substrate rotator and other high vacuum measuring gauges are also provided to enable systematic monitoring of deposition process [11]. Prior to deposition Si (100) substrates were cleaned by RCA-1 process [14].

Deposition of AlN-thin films was carried out at a base pressure of $4x10^{-6}$ mbar. Working pressure of the chamber was maintained at $2x10^{-4}$ mbar while substrate temperature was kept constant at $500^{o}$C. Films were grown for 3, 5, 8 and 15 minutes. During deposition, the metal atom flux was supplied by sputtering Al target with an inert ion beam of $Ar^+$ ions having energy of 500 eV. Reactive assisted flux of $N_2$ plasma was provided by an ion assisted source operated at energy of 100 eV.

Surface morphology of deposited films was analyzed using an Atomic Force Microscope (AFM) (Ntegra Prima of M/s NT-MDT, Russia) in semi-contact mode. Si-tip of radius 35 nm was used to scan 1x1 $\mu m^2$ area at several places over a 10mm X 10mm film surface. Root mean square of surface heights at different points on the film surface was taken as roughness value of the film. To improve the statistics and representivity, images and roughness line profiles were acquired at several points on the film surface.

Static scaling exponent "α" was calculated by log-log plot of interfacial width (ξ) and scan length (L) belonging to each of film. Dynamic scaling exponent "β" was obtained from the slope of the curve generated by log-log plot of rms surface roughness (δ) vs. deposition time (t). The autocorrelation function ($G(|r|)$) and power spectral density PSD expressed as g(q), was used to differentiate between roughening and smoothening phenomena occurring on the film surface. This approach is similar to the one used for our previous study [11].

## 3. Results and Discussions

Figure. 1 shows atomic force micrographs of AlN thin films grown by IBSD for different time durations of 3 minutes, 5minutes, 8 minutes and 15 minutes in reactive assistance of $N_2$ plasma. It is inferred well from these micrographs that low energy ion/surface interaction modifies microstructure of the film by altering surface phenomena prevailing at different growth stages. Figure 1(A) shows that small cone shaped nuclei nucleates at the substrate surface after a deposition duration of 3 minutes. It is clearly seen in the micrograph that number density of nuclei increases in the courses of reactive assistance of $N_2$ plasma. The nucleation process is found to be accelerated as compared to the case where there is no provision of reactive assistance [7]. With increase in deposition time to 5 minutes, the nuclei grew more rapidly in number than in size. Further assistance of $N_2$ plasma at high substrate temperature, accelerates the coalescence stage of films growth there by turning the film into a continuous one. This is observed after a deposition of 8 minutes. This further indicated that reactive assistance of $N_2$ plasma increased the number density of small cone shaped clusters substantially as compared to the situation where Al thin film was grown without any ion assistance [7]. A densely packed AlN thin film was encountered after deposition duration of 15 minutes.

Representative line profiles of the surfaces after laps of corresponding deposition times are shown in figure. 2. It is observed that height fluctuations and lateral aggregation among these cone shaped clusters are two pragmatic parameters to investigate the surface growth occurring over each deposition time. For a 3 minute deposition duration, tiny nuclei represented by small height amplitudes and spatial coverage appearing on the substrate surface were well separated from each other. As the deposition time was increased to 5 minutes, frequency of their appearance also increased and their number density increased more in pronounced manner compared to the spatial coverage. Height amplitudes and spatial coverages of the peaks were found to increase significantly for an 8 minute deposition duration while after 15 minutes of deposition, the film became continuous as all the peaks appearing in the corresponding line profile are found to average out.

According to dynamic scaling theory, static scaling exponent 'α' can be calculated by plotting log(ξ) vs. log(L) which is shown in figure 3 for each deposition time. It is observed from the graphs that ξ increases with increase in the length scale L and after a certain critical length $L_{c,}$ ξ gets saturated. This saturated value of $\xi_L$ represents the rms roughness 'δ' of the surface under investigation. In the early stages of film growth, $\alpha = 0.62$ governs the evolution of surface roughness upto a critical length of $L_c$ = 600 nm after which a saturated value of rms surface roughness $\delta = 1.99$ nm is achieved. This is shown in figure 3(A). As the deposition time is increased to 5 minutes, 'α' decreased significantly to 0.46 having same $L_c$ but $\delta$ increased to 2.15 nm as shown in figure 3(B). Figure 3(C) shows that for a deposition time of 8 minutes, '$\alpha$' further decreases to 0.36 along with a decrease in $L_c$ to 550 nm. This is despite the fact that the associated rms value of roughness '$\delta$' of the surface increases to 3 nm as assisting nitrogen ions get more time to interact with the surface where by more energy to the arriving flux of atoms is added which enhances surface diffusion. When deposition was carried out for longer duration of 15 minutes, evolution of the surface is governed by even smaller $\alpha = 0.32$ with a corresponding decrease in $L_c$ to 450 nm. But rms roughness '$\delta$' is found to increase further to 3.42 nm as a consequence of $N_2$ plasma assistance which transfers additional energy to surface and arriving atoms so that their mobility on the surface is enhanced. This enhanced mobility allows atoms to occupy different sites which causes superior coverage and producing denser film formation. Figure 4 shows how static scaling exponent '$\alpha$' decreases as the deposition time is increased. Dynamic scaling exponent '$\beta$' of the film is calculated from the slope of $\log(\delta)$ vs. $\log(t)$ curve shown in figure 5. It is clearly observed from the graph that there is a gradual increase in rms value of roughness $\delta$ of the film which is governed by $\beta = 0.36$.

Height correlation across lateral direction on all the samples was predicted by autocovariance function $G(|r|)$ as shown in figure 6. Figure (7) is the Fourier transform of the equation (2) which represents power spectral density $g(q)$ or PSD. To identify the controlling step from a host of material surface related phenomenological events like plastic flow, evaporation-condensation, bulk and surface diffusions which govern the growth of AlN thin films to different extents at each deposition time, $g(q)$ given in equation (3) was fitted with equation (4). $n = 1,2,3,4$ in equation (4) represent above phenomenona usually constitutes dominant mode of thin film growth at a given set of process conditions.

Figure 8 shows the fitting of experimentally obtained g(q) in equation (3) with equation (4). Sign of the parameters $\xi_n$ as shown in table.2, decides whether a given phenomenoa corresponding to a particular n-value will be dominant either as smoothening or a roughening phenomenon. The fits for all four data sets in figure 8 consistently indicate that the phenomenon corresponding to n = 2 and n = 4 are primary roughening processes. The surface diffusion (n = 4) is the dominating one while evaporation and recondensation (n = 2) also contribute significantly to the evolution of surface morphology. Phenomena corresponding to n = 1 and n = 3 are primary smoothening processes where bulk diffusion (n = 3) dominates over plastic flow (n = 1). Nonetheless, these aspects do contribute to evolution of surface morphology.

## 4. Conclusion

Reactive dual ion beam sputter deposition of AlN thin films was carried out for the analysis of surface growth characteristics by Atomic Force Microscopy. The variation of roughness as a function of deposition time was analysed by Dynamic Scaling Theory (DST). Two distinct exponents (static and dynamic) were used to unravel the film growth characteristics.

Following inferences were arrived from the studies:

i) Small cone shaped nuclei and clusters of AlN can be grown for 3 minutes and 5 minutes depositions, respectively, under the same deposition parameters.

ii). As the deposition time increased, static scaling exponent ‘α’ decreased gradually and substrate surface coverage was increased which is indicated by a decrease in critical length $L_{c.}$

iii). The rms roughness of the film was increased from 1.99 to 3.42 nm as the deposition time was increased from 3 minutes to 5 minutes. This is represented by Dynamic scaling exponent ‘β’ which was calculated from the slope of log(δ) vs. log(t) curve shown in figure 5.

iv). Surface diffusion (n = 4) becomes the major roughening phenomenon while evaporation recondensation (n = 2) also contributes significantly towards the evolution of surface morphology.

v). Bulk diffusion (n = 3) turns into the dominating smoothening phenomenon while plastic flow (n = 1) also contributes to smoothen the surface.

**Acknowledgement**

The authors would like to thank Mr. M. P. Janawadkar, Director, material Science Group for his encouragement and support.

**References:**

[1] P.J. Martin, H.A. Macleod, R.P. Netterfield, C.G. Pacey, W.G. Sainty, Appl. Opt. 22/1 (1983) 178.
[2] A.A. Galuska, Nuclear Instruments and Methods in Physics Research Section B: Beam Interactions with Materials and Atoms 59–60, Part 1/0 (1991) 487.
[3] J. Freisinger, J. Heland, D. Krämer, H. Löb, A. Scharmann, Review of Scientific Instruments 63/4 (1992) 2571.
[4] Y.-Y. Chen, J.-C. Hsu, P.W. Wang, Y.-W. Pai, C.-Y. Wu, Y.-H. Lin, Applied Surface Science 257/8 (2011) 3446.
[5] J.S. Colligon, Philosophical Transactions of the Royal Society of London. Series A: Mathematical, Physical and Engineering Sciences 362/1814 (2004) 103.
[6] M. Rusanen, I. Koponen, J. Heinonen, J. Sillanpää, Nuclear Instruments and Methods in Physics Research Section B: Beam Interactions with Materials and Atoms 148/1–4 (1999) 116.
[7] K.H. Müller, Journal of Applied Physics 59/8 (1986) 2803.
[8] S. A. L. Barbasi, H. E., Fractal Concepts in surface growth, 1995.
[9] G. Zhang, B.L. Weeks, M. Holtz, Surface Science 605/3–4 (2011) 463.

[10] F. Family, Journal of Physics A: Mathematical and General 19/8 (1986) L441.
[11] N. Sharma, Thin Solid Films (2014).
[12] W. Tong, R. Williams, Annual Review of Physical Chemistry 45/1 (1994) 401.
[13] C. Herring, Journal of Applied Physics 21/4 (1950) 301.
[14] W. Kern, Journal of The Electrochemical Society 137/6 (1990) 1887.

**Figure.1**

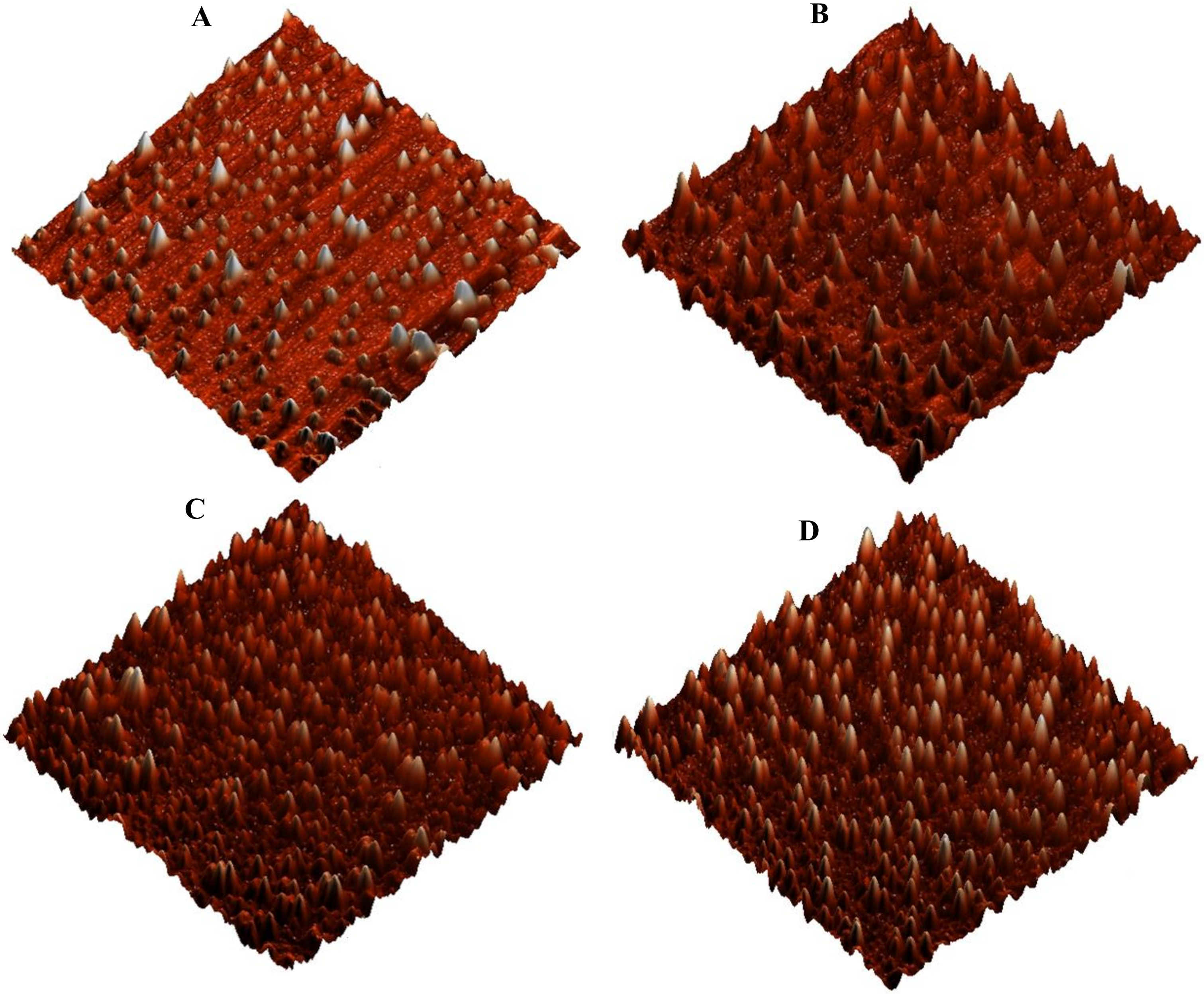

Representative 3D AFM images (1 X 1 $\mu m^2$) of the evolution of surface morphology for the films grown for different deposition times as, (A) 3 minutes, (B) 5 minutes, (C) 8 minutes and (D) 15 minutes.

**Figure.2**

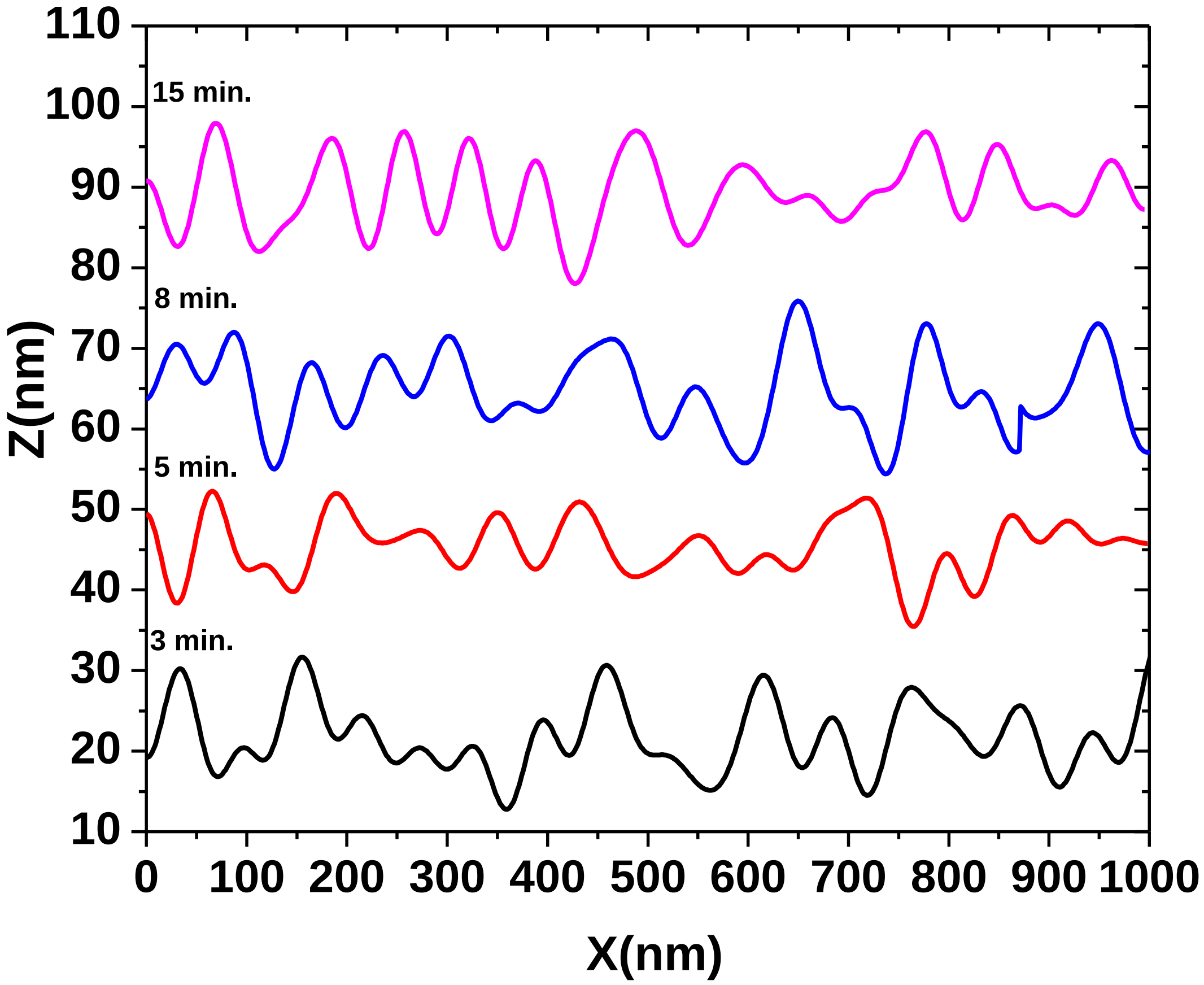

Representative line profiles across the sample surface for different deposition times

**Figure. 3**

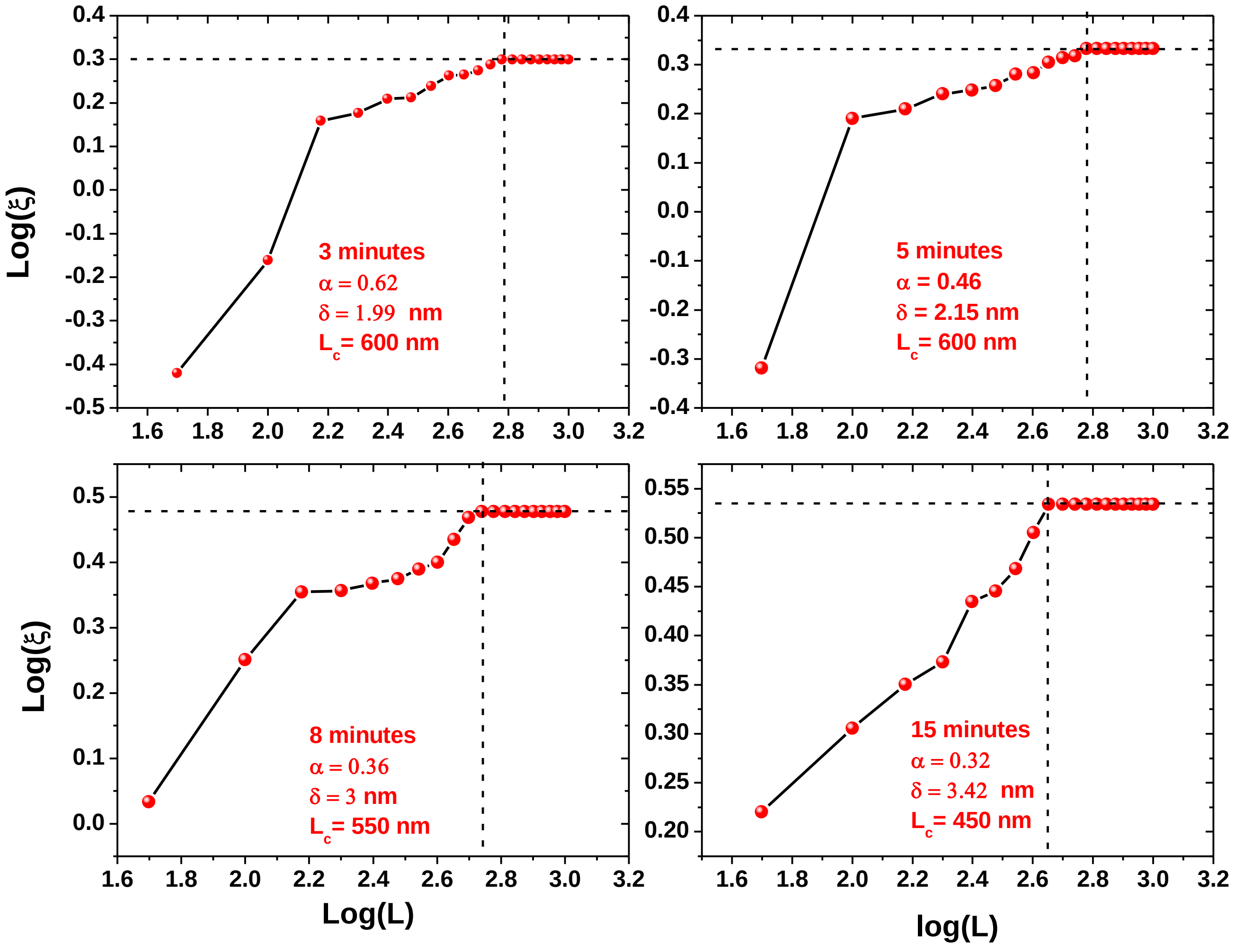

Plot of the interface width of the substrate surface as a function of length scale for several different growth times varied as (A) 3 minutes (B) 5 minutes (C) 8 minutes and (D) 15 minutes.

**Figure. 4**

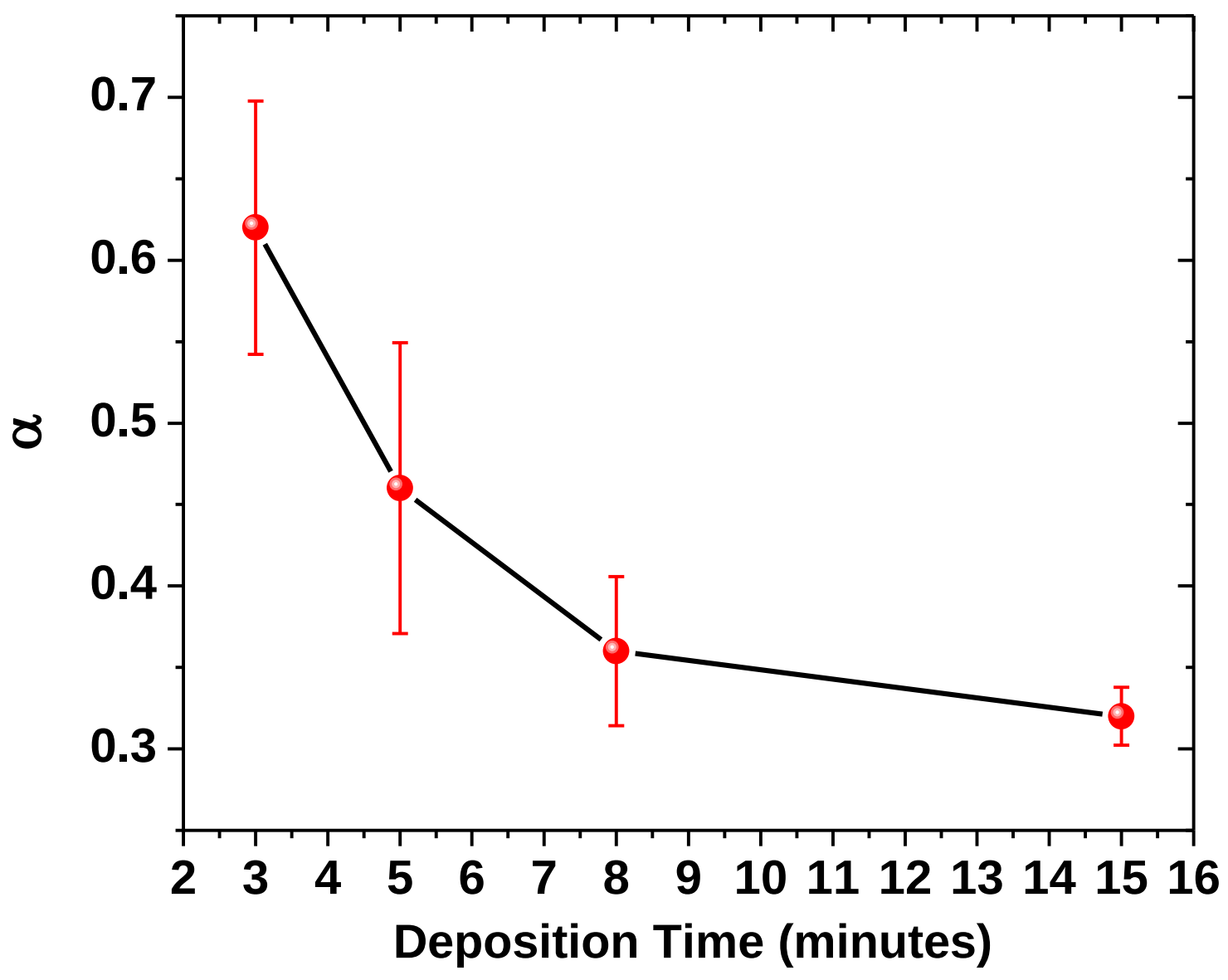

Variation of α–exponents obtained for different deposition times.

**Figure. 5**

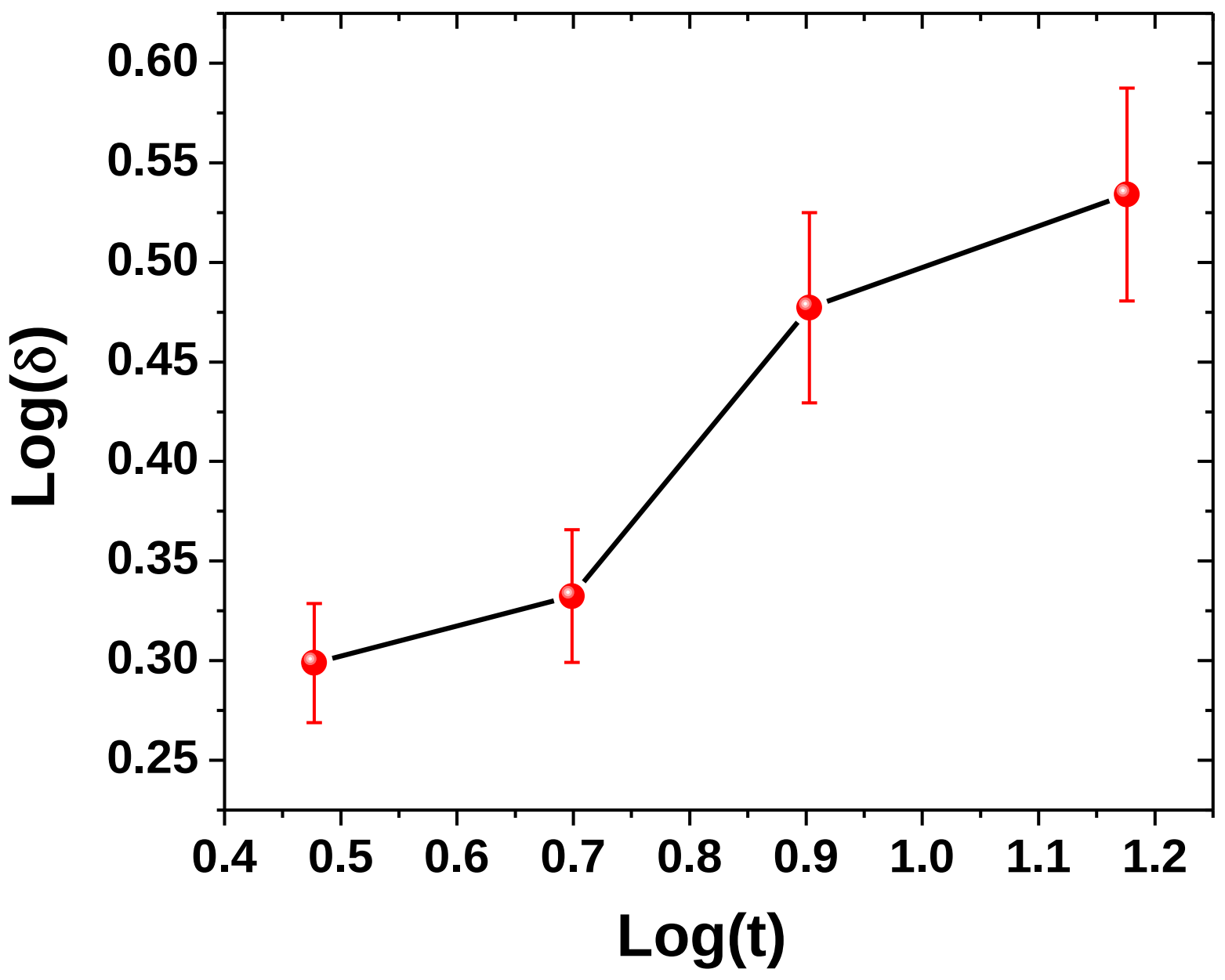

Variation of δ obtained for different deposition times

**Figure.6**

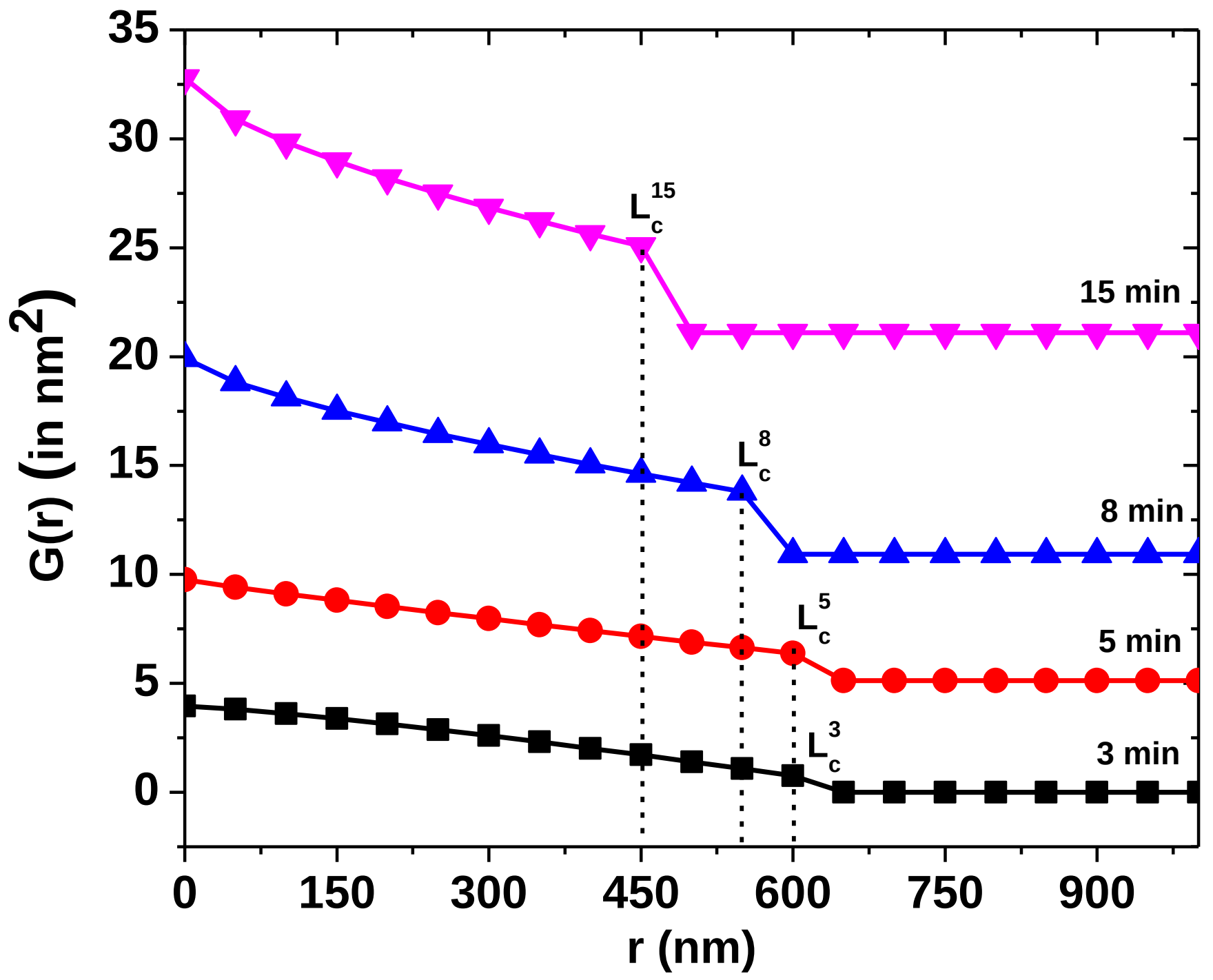

Autocovariance function for different deposition times depicting how heights are co-related at different points across lateral direction

**Figure. 7**

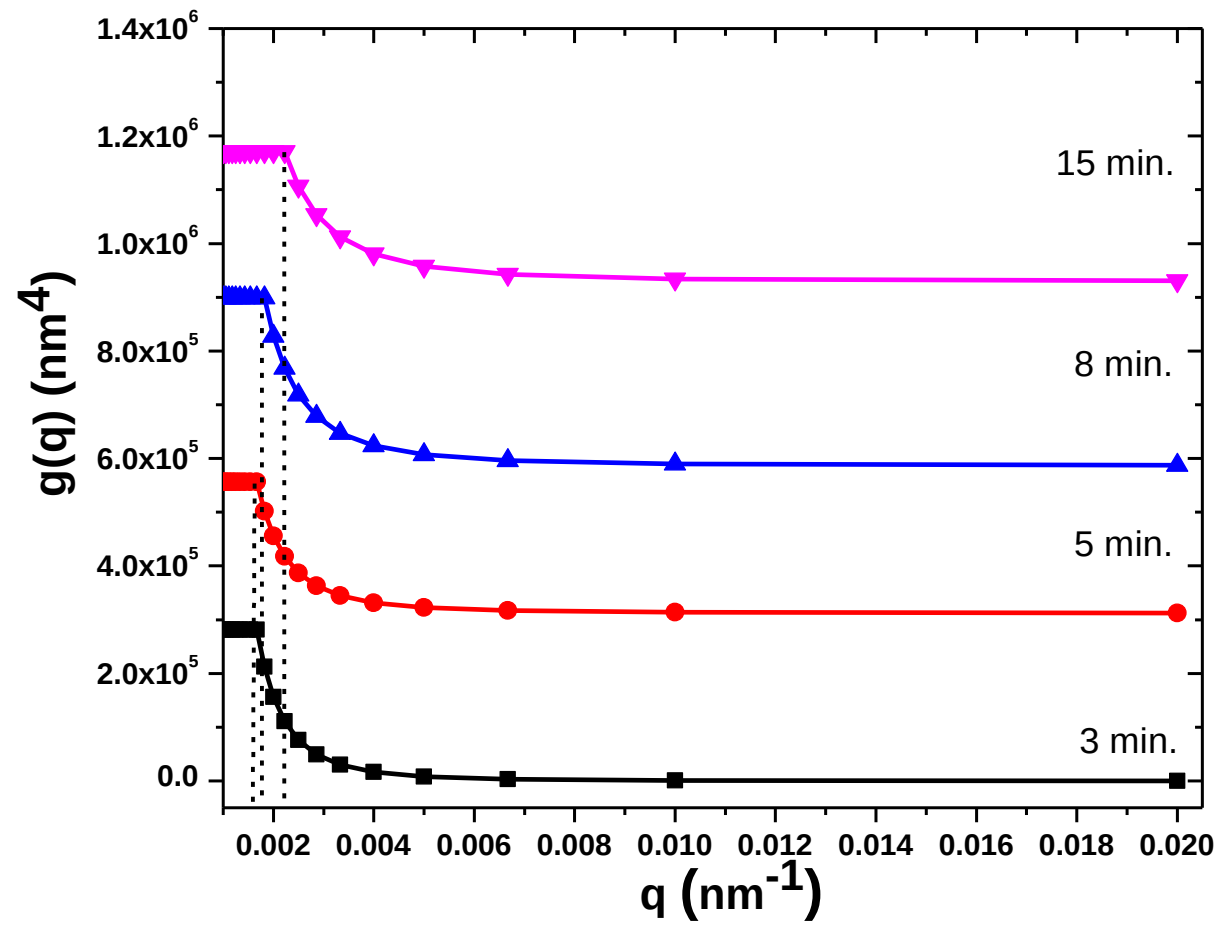

Spectral power density obtained corresponding to different deposition times an verifying critical lengths obtained for each deposition time.

## Figure. 8

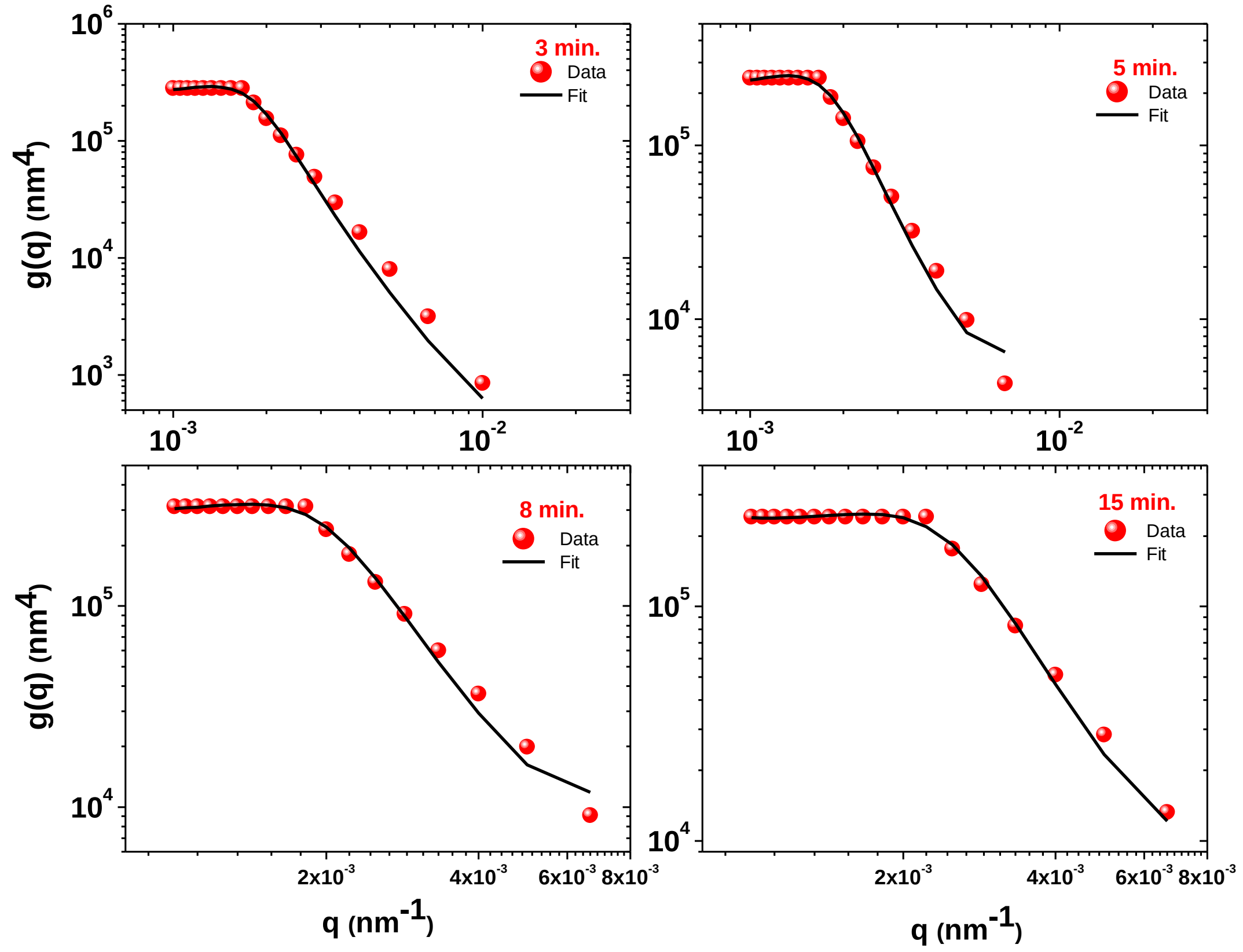

Fitting of experimentally obtained g(q) with equation (7) for different deposition times indicating towards dominating roughening/smoothening phenomenons. The data are presented as log–log plots where (A) 3 minutes, (B) 5 minutes, (C) 8 minutes and (D) 15 minutes.

## Table-1

| Deposition Time 't' (in min.) | α | $L_C$ (in nm) | δ (in nm) |
|---|---|---|---|
| 3 | 0.62 | 600 | 1.99 |
| 5 | 0.46 | 600 | 2.15 |
| 8 | 0.36 | 550 | 3.0 |
| 15 | 0.32 | 450 | 3.42 |

## Table-2

| Deposition Time (minutes) | Ω (X $10^5$) | $\xi_1$(X $10^3$) | $\xi_2$(X $10^6$) | $\xi_3$($10^8$) | $\xi_4$(X $10^{10}$) |
|---|---|---|---|---|---|
| 3 | 1.6 | -1.9 | 2.1 | -7.6 | 3.1 |
| 5 | 1.5 | -2.2 | 2.6 | -10.3 | 9.2 |
| 8 | 2.1 | -2.1 | 2.2 | -8.1 | 7.0 |
| 15 | 1.8 | -1.8 | 1.5 | -4.6 | 3.3 |